\documentclass[letter]{aa}
\usepackage{psfig}
\begin{document}
   \headnote{Letter to the Editor}

\title{The distance to the Pleiades from orbital solution of the double-lined
       eclipsing binary HD 23642\thanks{The photometric and spectroscopic
       observations discussed in this paper are electronically available
       from the web page
       http://ulisse.pd.astro.it/HD\_23642/}$^,$\thanks{Based in part on spectra
       collected with the ELODIE spectrograph at the 1.93m telescope of the
       Observatoire de Haute Provence (OHP), France}
       }

\author{
       U. Munari\inst{1}
       \and
       S. Dallaporta\inst{2}
       \and
       A. Siviero\inst{1,3}
       \and
       C. Soubiran\inst{4}
       \and
       M. Fiorucci\inst{1}
       \and
       P. Girard\inst{4}
       }

\offprints{U.Munari}

\institute {
INAF Osservatorio Astronomico di Padova, Sede di Asiago, I-36012 Asiago (VI), Italy
\and
Via Filzi 9, I-38034 Cembra (TN), Italy
\and
Astrophysical Observatory of the Dept. of Astronomy, Univ. of Padova,  
I-36012 Asiago (VI), Italy
\and
Observatoire Aquitain des Sciences de l'Univers, CNRS UMR 5804,
BP 89, 33270 Floirac, France
           }

\date{Received date..............; accepted date................}

\abstract{Combining precise $B$,$V$ photometry and radial velocities, we have
been able to derive a firm orbital solution and accurate physical parameters
for the newly discovered eclipsing binary HD~23642 in the Pleiades open
cluster. The resulting distance to the binary and therefore to the cluster
is 132$\pm$2~pc. This closely confirms the distance modulus obtained by
classical main sequence fitting methods ($m - M$=5.60 or 132 pc), moving
cluster techniques and the astrometric orbit of Atlas. This is the first
time the distance to a member of the Pleiades is derived by orbital solution
of a double-lined eclipsing binary, and it is intended to contribute to the
ongoing discussion about the discordant Hipparcos distance to the cluster.

         }

\authorrunning{U.Munari et al.}
\titlerunning{The distance to the Pleiades from the eclipsing binary HD 23642}
\maketitle

\section{Introduction}

The distance to the Pleiades has been derived by main-sequence fitting
methods several times over the years, and consistently found to cluster
around $m - M$=5.60~$\pm$0.04 (e.g. Turner 1979, Meynet et al. 1993),
corresponding to a distance of 132~($\pm$2) pc. Values of $m - M$=5.52
(127~pc) by Mitchell \& Johnson (1957) and $m - M$=5.75 (141~pc) by Eggen
(1950) bracket the range of published distance moduli.
Therefore, it came as a surprise the shorter distance of
116~$\pm$3.3~pc ($m - M$=5.33~$\pm$0.06) derived from the parallaxes of
54 Pleiades members when the results of the Hipparcos astrometric mission
became available (van Leeuwen \& Hansen Ruiz 1997). The $\sim$0.3~mag
difference has far reaching consequences in many areas of astrophysics, 
and it soon prompted extensive observational and theoretical work to account for it.

     \begin{figure*}[!t]
     \centerline{\psfig{file=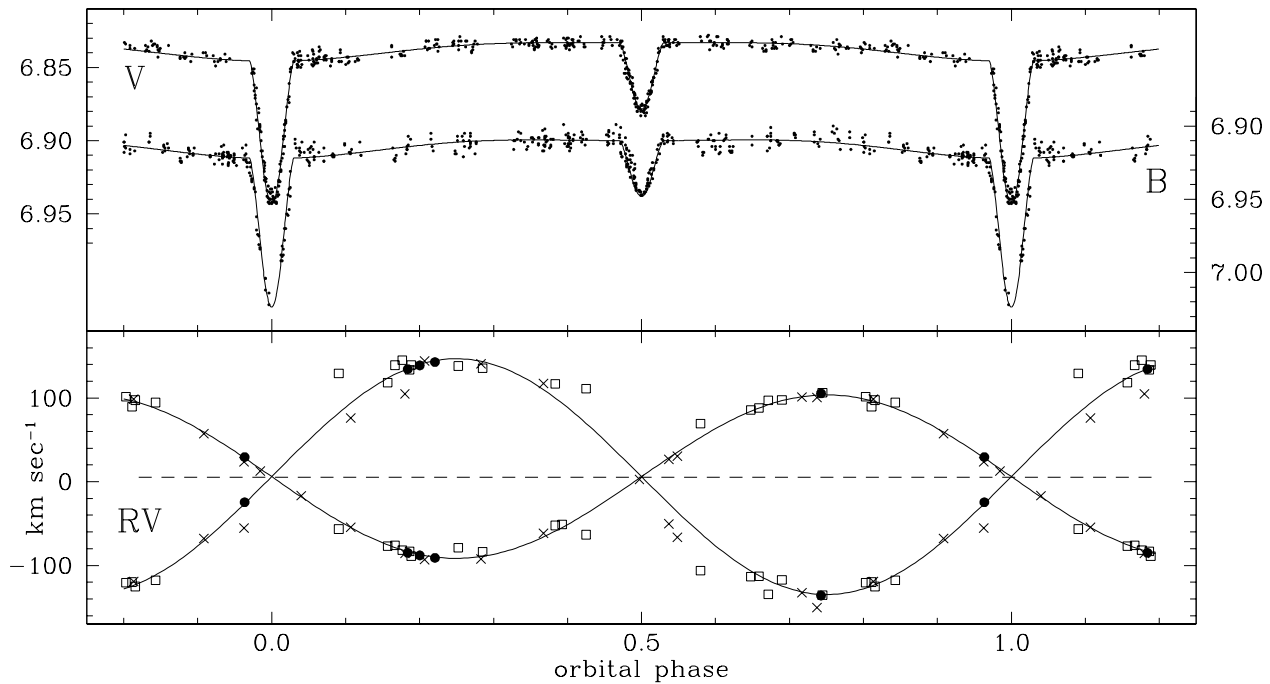,width=16.7 cm,angle=270}}
     \caption[]{Our photo-electric $V$ (top) and $B$ (bottom) observations of
                HD~23642 are phase plotted in the upper panel following the
                ephemeris of Table~2.  The lower panel displays our radial
                velocities from Table~1 (filled circles) and those from
                Pearce (1957, squares) and Abt (1958, crosses).  The curves
                over plot the orbital solution given in Table~2.}
     \end{figure*}

An anomalous abundance of Helium was discussed as a possible explanation of
the large difference between Hipparcos and ground-based photometric
distances to the Pleiades (Mermilliod et al. 1997a). The required helium
over-abundance ($Y$=0.37) is however too large to be a feasible explanation
according to Pinsonneault et al. (1998). A sub-solar metallicity could
reconcile the distances, but the metallicity of the Pleiades has been
measured several times and with different methods, which generally converged
to a mean value of [Fe/H]=0.00$\pm$0.03 (e.g. Stauffer et al. 2003).
Castellani et al. (2002) were able to obtain a good fit of the Pleiades main
sequence with their theoretical isochrones (with updated physical inputs)
and the Hipparcos distance, employing a sub-solar metallicity of
[Fe/H]=--0.15. Working differentially with a sample of nearby G and K stars,
Percival et al. (2003) presented a re-calibration of photometric metallicity
indexes and argued that they support a sub-solar metallicity of the Pleiades
in spite of the spectroscopic evidences (e.g. Boesgaard \& Friel 1990) for a
solar one, therefore removing the difference between main-sequence fitting
and Hipparcos distances. However, Stauffer et al. (2003) remarked about the
anomalous blue colors of K stars in the Pleiades, that they ascribed to fast
rotation and spotted surfaces as a consequence of the young age of the
cluster.

Narayanan \& Gould (1999) derived a distance modulus to the Pleiades of $m
- M$=5.58~($\pm$0.18) using a variant of the moving cluster method, the
gradient in the radial velocity of the cluster members in the direction of
the proper motion of the cluster. Their techniques relies on the assumption
that the velocity structure of the Pleiades is not significantly affected by
rotation. Narayanan \& Gould concluded that the errors in the Hipparcos
parallaxes toward Pleiades are spatially correlated over angular scales of a
few degrees, with an amplitude up to 2~mas. Almost simultaneously, however,
van Leeuwen (1999) compared the results on 9 clusters and concluded instead
that the Hipparcos parallaxes of the Pleiades were basically unaffected by
systematics and that the distance modulus to the latter is $m - M$=5.37~($\pm$0.07).
Makarov (2002) suggested that, while Hipparcos parallaxes are overall of
excellent quality, those for the Pleiades suffered from data analysis
procedures that could have introduced systematics when a rich and bright cluster
was crossing one of the two Hipparcos fields of view while the other was
essentially deprived of stars, as in the case of Pleiades. From 
intermediate Hipparcos data and under some assumptions, Makarov has
recomputed the distance to the Pleiades as 129~$\pm$3~pc. F. van Leeuwen (private
communication) is currently working on a further iteration on original 
Hipparcos reductions that should assess and quantify is any systematics
affect the Pleiades data.

The question about the distance to the Pleiades has not yet been firmly
solved, and to tackle it new, robust and independent approaches (as much
geometric as possible) are required. A first one, advocated by Pacz\'{y}nski
(2003), combines astrometric and spectroscopic observations of Atlas
(HD~23850), one of the brightest members of the Pleiades, which is an
astrometric binary with a period of 291~day, a semi-major axis of 12.9~mas
and 0.246 eccentricity. Pan et al. (2004) did not have the radial
velocities, but they anyway derived the distance by combining the
astrometric orbit with a mass-luminosity relation. They derived a distance
of 135$\pm$2~pc, which is in close agreement with the results of
main-sequence fitting methods.

Another method involves double-lined eclipsing binaries (SB2 EB), which are
the distance indicators now providing the most reliable distances to
Magellanic Clouds and other galaxies in the Local Group. The recent
discovery by Torres (2003) that HD~23642 in the Pleiades is an SB2 eclipsing
binary (the only one so far known in the cluster) prompted us to use it to
measure the distance to the Pleiades. In this {\em Letter} we present
accurate new $B$ and $V$ photoelectric photometry and radial velocities of
HD~23642, and use them to derive an orbital solution and infer the distance
to the star, and therefore to the cluster.

\section{Observations}

We observed HD~23642 in $B$ and $V$ (standard Johnson filters) from a private
observatory near Cembra (Trento), Italy. The instrument was a 28 cm
Schmidt-Cassegrain telescope equipped with an Optec SSP5 photometer. 
It proved already to be a very accurate and reliable instrument (cf. Siviero 
et al. 2004) perfectly suited to deal with the low amplitude of HD~23642 eclipses
($\sim$0.1 mag). HD\,23568 (HIP\,17664, $V_T$=6.824$\pm$0.011, $B_T= 6.842\pm0.015$,
spectrum B9.5V) was chosen as comparison star and HD\,23763 (HIP\,17791,
$V_T$=6.963$\pm$0.011, $B_T=7.109\pm 0.016$, spectrum A1V) as a check star.
Following the Bessell (2000) transformations between Tycho and Johnson
photometric systems, we adopted $V$=6.830 and $B$=6.851 for the comparison
star. The comparison star was measured against the check star at least once
every observing run, and found constant with standard deviations of 0.005 mag 
in $V$ and 0.006 mag in $B$, confirming the Hipparcos photometric results.
In all, 492 measurements in $V$, and 432 in $B$ were collected of HD~23642
between Aug. 19, 2003 and Feb. 16, 2004. All the observations were corrected
for atmospheric extinction and instrumental color equations (via calibration
on Landolt's equatorial fields), and the instrumental differential
magnitudes were transformed into the standard Johnson UBV system. The
variable, comparison and check stars are similar in spectral type, lie very
close on the sky (15 arcmin) and were always observed at zenith distances
$<60^\circ$, which argues for a high consistency of our photometry.

High resolution spectra of HD~23642 were obtained at five distinct epochs
with the ELODIE echelle spectrograph at the 1.93m telescope of the
Haute-Provence Observatory (Baranne et al. 1996). ELODIE covers the spectral
range 3900-6800~\AA\ in a single exposure as 67 orders at a mean resolving
power of 42000. Optimal extraction and wavelength calibration were performed
with the on-line automatic reduction pipeline. Radial velocities for the two
components are reported in Table~1. They were measured by both classical
cross-correlation against suitable A-type templates and by the Least-Square
Deconvolution method of Donati et al (1997). The errors given in Table~1 
reflect the scatter of the measurements obtained with the two methods run
with different parameters.

\section{The temperature of the primary and the reddening}

     \begin{table}[!t]
     \caption{Heliocentric radial velocities of HD~23642. The S/N is evaluated
     around 5500~\AA.}
     \centerline{\psfig{file=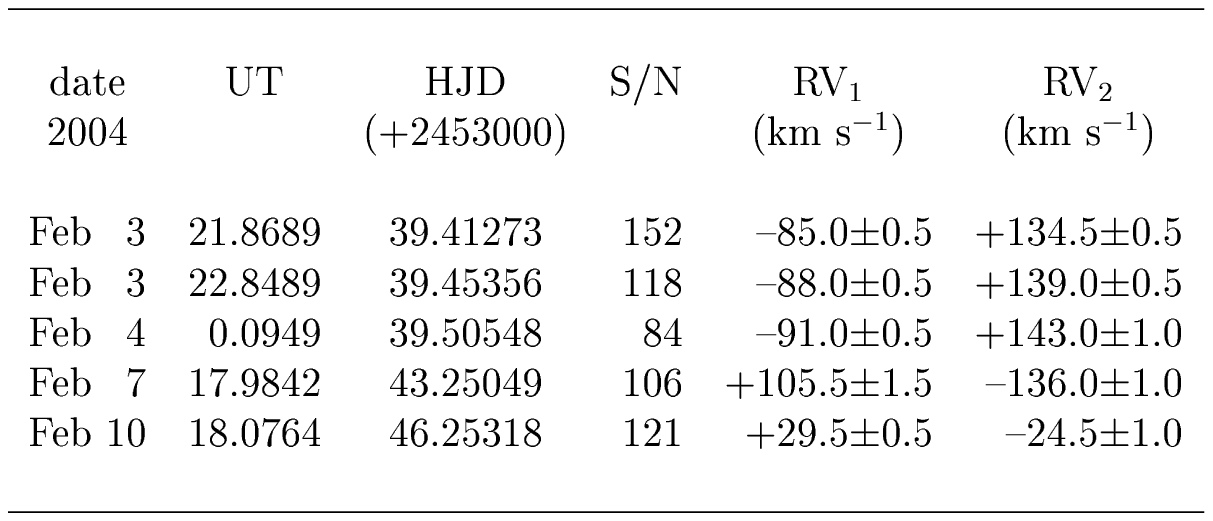,width=8.5cm}}
     \end{table}

     \begin{table}[!t]
     \caption{Orbital solution for HD~23642 (over plotted to observed data in
     Figure~1).  Formal errors of the solution are given. $\dag$: derived
     from analysis of multi-band photometry and fixed.}
     \centerline{\psfig{file=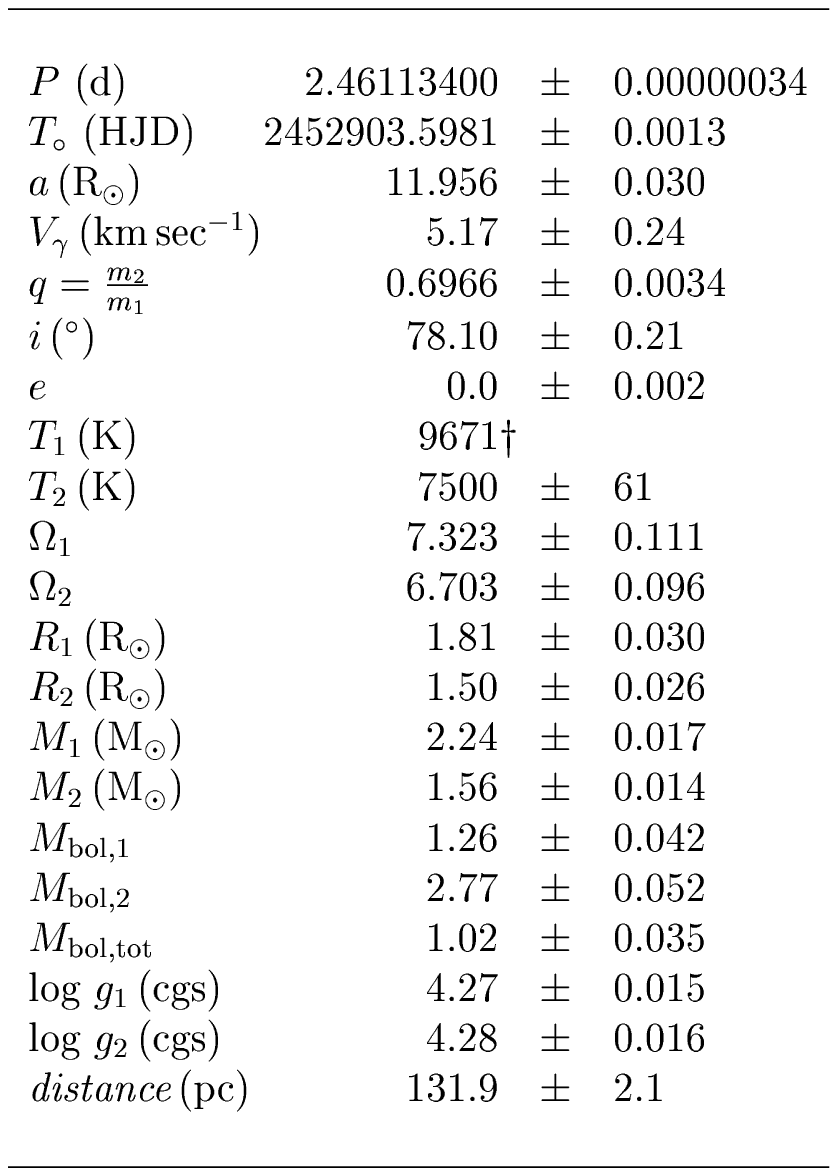,width=6.5cm}}
     \end{table}

One parameter, the temperature of the primary ($T_1$), cannot be directly
modeled in SB2 EB analysis and must be determined independently. In the case
of HD~23642, the spectral classification reported in literature is too
sparse, ranging from B9 to A1. To determine $T_1$ we then turned to
photometry and searched the literature and the General Catalogue of
Photometric Data (GCPD, Mermilliod et al. 1997b) for observations of
HD~23642. Data were found in the Walraven~{\em VBLUW}, Str\"{o}mgren~{\em
ubvy}, {\em WBVR}, Geneva, DDO, Vilnius, Johnson~{\em UBVRI}, Tycho, 2MASS
and ASN~74 photometric systems (Crawford \& Perry 1976, Kornilov et al.
1991, Morel \& Magnenat 1978, Persinger \& Castelaz 1990, van Leeuwen et
al. 1986, Wesselius et al. 1982, GCPD).  Using the relevant parameters for
these systems as censed in the Asiago Database on Photometric Systems (Moro
\& Munari 2000, Fiorucci \& Munari 2003), they were applied to the
extensive synthetic spectral atlas of Munari et al. (2004). A total of 2310
synthetic spectra of the binary were retrieved from it considering all
possible combinations of 7 different temperatures for the primary (9000$\leq
T_2 \leq$10500~K), 10 for the secondary (6750$\leq T_2 \leq$9000~K), 3
values for the gravity ($\log g$=4.0, 4.5, 5.0) and 11 for the reddening
(0.00$\leq E_{B-V} \leq$0.10). The [Fe/H]=0.00$\pm$0.02 metallicity of the
Pleiades reported by Stauffer et al. (2003) was adopted, as well as the
standard $R_V=A_V/E_{B-V}$=3.1 reddening law (Fitzpatrick 1999) that was
found to be valid for the Pleiades by Guthrie (1987). A minimum distance
method was then applied to observed and computed colors in each photometric
system, which provided pretty consistent results among them. The weight
averaged means for the temperature of the primary and the reddening were
well constrained to $T_1 = 9671\pm46$~K and $E_{B-V}$=0.012$\pm$0.004. Less
constrained were the temperature of the fainter secondary ($T_2 =
8023\pm544$~K) and the surface gravity of the primary ($\log
g_1$=4.17$\pm$0.13), while the gravity of the secondary gave no measurable
signal.

The value of the reddening is perfectly confirmed by the equivalent width of
the interstellar NaI~D lines. We measured them on our high resolution
spectra and found E.W.(NaI 5890~\AA)=0.035$\pm$0.001~\AA\ that corresponds
to $E_{B-V}$=0.011 adopting the Munari \& Zwitter (1997) calibration. White
et al. (2001) has derived a closely similar equivalent width
(0.036$\pm$0.003~\AA) from a single, higher resolution spectrum.

\section{Orbital solution and distance to the Pleiades}

To augment the number and phase coverage of the radial velocities, in the
orbital solution we also considered the measurements by Pearce (1957) and
Abt (1958). The adopted average relative weights are 10, 2, 1.5 for ours,
Pearce and Abt data, respectively.

The orbital solution of HD~23642 has been obtained with version WD98K93d of
the Wilson-Devinney code (Wilson \& Devinney 1971) as modified by Milone et
al. (1992) to include Kurucz's model atmospheres to approximate the surface
fluxes of the two stars. The computations have been run within {\tt Mode-2}
program option.  Limb darkening coefficients have been taken from Van Hamme
(1993) interpolated for the metallicity, temperature and gravity appropriate
for the components of HD~23642. A linear law for limb darkening and single
reflection (i.e. only the inverse square law illumination) were considered. The
bolometric albedos were set to 1.0 and tests on the lightcurve confirmed the
choice. Finally, a gravity brightening exponent $\beta$=1.0 has been adopted
consistent with the radiative nature of the atmospheres in HD~23642. The
effects on the orbital solution of the adopted limb darkening coefficients
and law, bolometric albedos, single reflection scenario and gravity
brightening exponent are essentially negligible. Their cumulative impact on
the final distance to HD~23642 does not exceed $\sim$1 pc. The orbital solution
is given in Table~2 and over plotted to the data in Figure~1. The parameters
in common with the radial-velocities-only solution of Torres (2003) are
very similar.

The two stars are well within their Roche lobes and not distorted. The orbit
appears relaxed to stationary conditions. In fact, the eccentricity is null
within a small uncertainty and the rotational velocity of the two
components, as measured on the high resolution spectra (37$\pm$1 for the
primary and 31$\pm$2~km~s$^{-1}$ for the secondary), is that expected for
the rotation period to be synchronous with the orbital period.

The barycentric radial velocity of HD~23642,
$V_\gamma$=5.17$\pm$0.24~km~s$^{-1}$, is coincident with the mean radial
velocity of the Pleiades within the velocity dispersion of its members. They
are $RV_{\rm cluster}$=5.19~km~s$^{-1}$ and $\sigma_{\rm
RV}$=0.70~km~s$^{-1}$ according to Smith \& Struve (1944), or $RV_{\rm
cluster}$=5.74$\pm$0.07~km~s$^{-1}$ and $\sigma_{\rm
RV}$=0.69$\pm$0.05~km~s$^{-1}$ following Narayanan \& Gould (1999).

The distance to HD~23642 resulting from our orbital solution is 132$\pm$2~pc
(cf. Table~2). This result does not depend on the measured reddening
for HD~23642. Rising it to $E_{B-V}$=0.035 average reddening for cluster
members (e.g. Crawford \& Perry 1976), would require an increase to
9910$\pm$145~K for the primary to fit observed colors (the larger error
indicates a much poorer $\chi^2$ fit to literature photometry compared to
$E_{B-V}$=0.012). The larger intrinsic brightness compensates for the extra
$\Delta$A$_V$=0.07~mag extinction and the resulting distance is 130.6$\pm$3.7~pc.
Even if the distance to HD~23642 derived in this paper concerns just one
cluster member, nevertheless it is in excellent agreement with the results
from the main-sequence fitting, moving cluster method and astrometric orbit
of Atlas. These results agree so well that the issue of the
distance to the Pleiades seems now addressed, and finally over.

\acknowledgements{We would like to thank M.A.C. Perryman and F. van Leeuwen
for reading the manuscript and commenting in detail on it. This work has been
supported in part by an Italian COFIN-2001 grant.}

\end{document}